\newcounter{myctr}

\documentclass{ws-acs}
\usepackage{cite}
\usepackage{graphics}
\begin{document}

\markboth{A. Johansson, D. Helbing and P. K. Shukla}{Evolutionary Model Specification Via Video Tracking} 

%
\catchline{}{}{}{}{}
%

\title{Specification of a Microscopic Pedestrian Model by
Evolutionary Adjustment to Video Tracking Data}

\author{\footnotesize Anders Johansson,$^1$ Dirk Helbing,$^{1,2}$ and Pradyumn K. Shukla$^{1,3}$}
\address{$^1$ Institute for Transport \& Economics, Dresden University of Technology,\\
Andreas-Schubert-Str. 23, 01062 Dresden, Germany\\[2mm]
$^2$ Collegium Budapest~-- Institute for Advanced Study,\\
Szenth\'{a}roms\'{a}g u. 2, 1014 Budapest, Hungary\\[2mm]
$^3$ Institute of Numerical Mathematics, Department of Mathematics, Dresden Universiy of Technology. 01062 Dresden, Germany}

\maketitle

\begin{history}
\end{history}

\begin{abstract}
Based on suitable video recordings of interactive pedestrian motion and improved tracking software,
we apply an evolutionary optimization algorithm to determine optimal parameter specifications for 
the social force model. The calibrated model is then used for large-scale pedestrian
simulations of evacuation scenarios, pilgrimage, and urban environments.
\end{abstract} 

\keywords{Evolutionary optimization; pedestrian interaction; video tracking; social force model;
pedestrian simulation; evacuation scenarios} 

\section{Introduction}

In recent years, pedestrian research has become a very active field. Queueing-theoretical
approaches (see, e.g., Refs.~\cite{mayne,lovas}) and fluid-dynamic models 
\cite{HelbingComplexSystems,Hughes2002,Hughes2003} have facilitated 
insights into the large-scale dynamics of pedestrian crowds, in particular into the
phenomon of intermittent flows and stop-and-go traffic emerging at extreme densities
\cite{HelbingJohanssonunddieDaenenPRL}. However, most recent work is dedicated to
microscopic simulation models such as the social force model \cite{Verkehrsdynamik,transci,panic,Hoog2002b,YuPRE},
cellular automata \cite{KirchnerEtAl,KlMeyeSc2001,Schadschneider2001}, particle hopping models
\cite{Nagatani}, and others, e.g. multi-agent approaches \cite{Willis2000,Batty2}.
\par
While some of these models are aiming at the simulation of real evacuation scenarios, others are
focussing on an understanding of the self-organized dynamical patterns that a realistic model
must also be able to reproduce. But how to calibrate these models? One common method is to 
modify model parameters and model components long enough until the simulations fit the
fundamental flow-density diagram such as the one proposed by Weidmann \cite{Weidmann}:
\begin{equation}
 V(\rho) = 1.34\mbox{ms}^{-1} 
 \left\{ 1 - \exp \left[ -1.913\mbox{m}^{-2} \left(\frac{1}{\rho} - \frac{1}{5.4\mbox{m}^2} \right)\right]\right\}\, .
\label{Weidmann}
\end{equation}
This method can certainly be applied, but it does not necessarily lead to realistic and reliable simulation results. First of all,
the fundamental diagram depends enormously on the body size distribution \cite{Teknomo} and other characteristic
features of the crowd 
Second, most models fitting the fundamental diagram 
must be separately calibrated to many differen situations, as they do
not reproduce the various dynamical phenomena in pedestrian crowds well. 
These phenomena include the formation
of lanes of uniform walking direction in crowds with oppositely moving pedestrians, stripe formation
in two intersecting pedestrian streams, and oscillations of the passing direction at bottlenecks
under normal conditions with moderate densities \cite{HelbinginEnvironmentandPlanningB28,transci}. 
\par
Apart from the latter two phenomena, which have been discovered only recently, 
the social force model, despite its simplicity, has been shown to reproduce the dynamics 
in crowds qualitatively well. However, not much effort has been spent in the past to calibrate it
for the application to dimensioning and evacuation problems. We will, therefore, introduce a
method to calibrate microscopic pedestrian simulation models using trajectory data 
reflecting pedestrian interactions, and apply it to the social force model.
\par
Our paper is organized as follows: Section~\ref{Sec1} will shortly introduce the social force model and
some specifications of it. In Sec.~\ref{Sec2}, we will then introduce our method to
optimize parameters based on trajectory data. For this, we will shortly describe the video tracking method
applied to gain the trajectory data. Moreover, we will discuss how the data can be used to determine
the interaction forces themselves. Afterwards, in Section~\ref{Sec3} we will present some examples of
large-scale pedestrian and evacuation simulations. Finally, Sec.~\ref{Sec4} will summarize and conclude our
paper.
\section{Specifications of the Social Force Model}\label{Sec1}

The social force model can be underpinned with a social science model of behavioral changes proposed
by Lewin \cite{Lew51}. He has introduced the idea that behavioral changes 
were guided by so-called {\it social fields} or {\it social forces}. This idea has been 
put into mathematical form by Helbing \cite{Hel91} and applied to opinion formation \cite{MathSoc}, pedestrian
motion \cite{BehSci91}, and vehicle traffic \cite{TilchHel}. The social force model for pedestrians assumes that 
each individual $\alpha$ is trying to move in a desired direction $\vec{e}_{\alpha}$ with a 
desired speed $v_{\alpha}^0$, and that it adapts the actual velocity $\vec{v}_{\alpha}$ to the
desired one, $\vec{v}_\alpha^0 = v_\alpha^0 \vec{e}_\alpha$ within 
a certain relaxation time $\tau_{\alpha}$. The velocity $v_{\alpha}(t)=d\vec{r}_\alpha/dt$, i.e. the temporal change 
of the location $r_{\alpha}(t)$,  is itself assumed to change according to the acceleration equation
\begin{equation}
\frac{d\vec{v}_{\alpha}(t)}{dt}=\vec{f}_{\alpha}(t)+\vec{\xi}_{\alpha}(t) \, ,
\end{equation}
where $\vec{\xi}_\alpha(t)$ is a fluctuation term and $\vec{f}_{\alpha}(t)$ the systematic
part of the acceleration force of pedestrian $\alpha$ given by
\begin{equation}
\vec{f}_{\alpha}(t)=\frac{1}{\tau_{\alpha}}(v_{\alpha}^0\vec{e}_{\alpha}-\vec{v}_{\alpha})
+\sum_{\beta(\ne\alpha)}\vec{f}_{\alpha\beta}(t)+\sum_{i}\vec{f}_{\alpha i}(t) \, .
\end{equation}
The terms $\vec{f}_{\alpha\beta}(t)$, $\vec{f}_{\alpha i}(t)$ denote the repulsive forces
describing the attempts to keep a certain safety distance to other pedestrians $\beta$ and obstacles $i$.
The fluctuation term $\vec{\xi}_{\alpha}(t)$ reflects random behavioral variations arising 
from deliberate or accidental deviations from the average way of motion. The above equation 
are nonlinearly coupled Langevin equations and can be solved numerically using Euler's method. 
In very crowded situations, additional physical contact forces come into play \cite{panic}. 
\par
For the time being, we will assume a simplified interaction force of the form
\begin{equation}
 \vec{f}_{\alpha\beta}(t) = w\big(\varphi_{\alpha\beta}(t)\big) 
 \vec{g}\big(d_{\alpha\beta}(t)\big) \, ,
\end{equation}
where $\vec{d}_{\alpha\beta} = \vec{r}_\alpha - \vec{r}_\beta$ is the distance vector pointing from 
pedestrian $\beta$ to $\alpha$ and $\varphi_{\alpha\beta}$ the angle between the normalized
distance vector $\widehat{\vec{d}}_{\alpha\beta} = \vec{d}_{\alpha\beta}/
\|\vec{d}_{\alpha\beta}\|$ and the direction
$\vec{e}_\alpha$ 
of motion of pedestrian $\alpha$, i.e.  
$\cos(\varphi_{\alpha\beta}) = \vec{e}_\alpha 
\cdot \widehat{\vec{d}}_{\alpha\beta}$. It has, for example, been suggested to reflect by the function
\begin{equation}
w\big(\varphi_{\alpha\beta}(t)\big) 
=\left(\lambda_{\alpha}+(1-\lambda_{\alpha})\frac{1+\cos(\varphi_{\alpha\beta})}{2}\right) 
\label{w}
\end{equation}
that the reaction of pedestrians to what happens in front of them is much stronger than 
to what happens behind them. Here, $\lambda_\alpha$ with $0 \le \lambda_\alpha \le 1$
is a parameter which grows with the strength of interactions from behind. 
Later on, we will try to determine this angular dependence
from video tracking data. The distance dependence $g\big(d_{\alpha\beta}(t)\big)$ has been
specified in different ways. We will start with the {\bf circular specification}
of the interaction force, 
\begin{equation}
 \vec{g}(d_{\alpha\beta}) = A_\alpha \mbox{e}^{(R_\alpha + R_\beta - d_{\alpha\beta})/B_\alpha} 
\widehat{\vec{d}}_{\alpha\beta} \, ,
\label{g}
\end{equation}
where $R_\alpha$ and $R_\beta$ denote something like the radii of pedestrians $\alpha$ and
$\beta$, while $A_\alpha$ and $B_\alpha$ are parameters. $A_\alpha$ reflects the 
strength of interaction, while $B_\alpha$ corresponds to the interaction range. While the dependence
on $\alpha$ explicitly allows for a dependence of these parameters on the single individual,
we will assume $A_\alpha = A$ and $B_\alpha = B$ in the following. Otherwise, it would be hard
to collect enough data for parameter calibration.
   
\section{Evolutionary Model Specification Via Video Tracking}\label{Sec2}

Let us now turn to the question how the parameters $A$ and $B$ and the angular 
interaction strength $w(\varphi_{\alpha\beta})$ can be determined from video recordings. 

\subsection{Video Tracking Method}

\begin{figure} \begin{center}
	\includegraphics[width=0.5\textwidth]{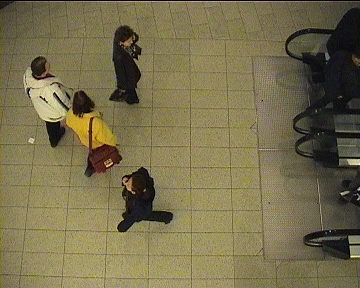}\\[2mm]
	\includegraphics[width=0.5\textwidth]{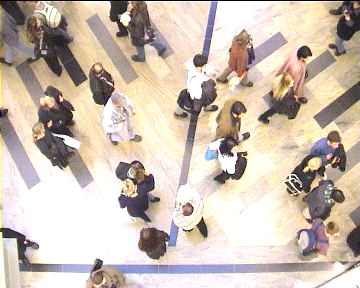}\\[2mm]
	\includegraphics[width=0.5\textwidth]{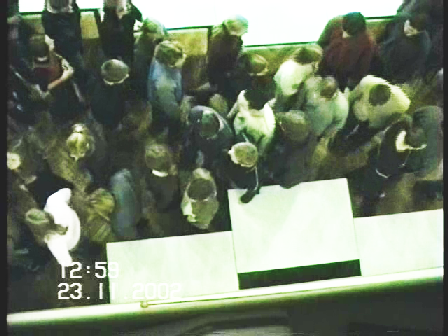}
	\caption{
Video recordings used for parameter calibration in
this study. Top: Entrance and exit area of two escalators. 
Middle: Free movement in a shopping mall in Budapest. 
Bottom: Pedestrian crossing experiment with students (see Ref.~\cite{transci} for details). At least
in the first two cases, pedestrians did not know they were recorded. Therefore, it
can be assumed that their behavior was not influenced and representative for middle European
conditions.}\label{Fig1}
\end{center} 
\end{figure}
\begin{figure}[htb] \begin{center}
    \includegraphics[width=0.4\textwidth]{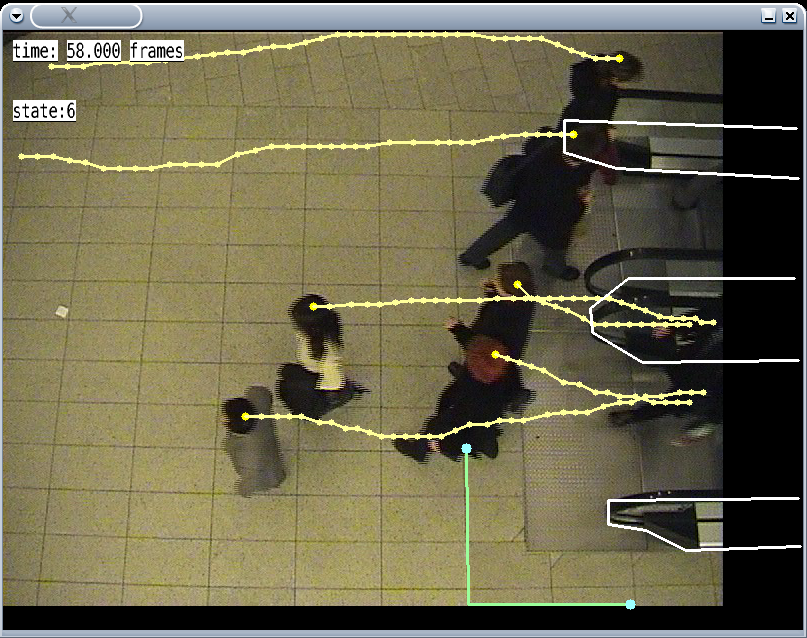}
    \includegraphics[width=0.45\textwidth]{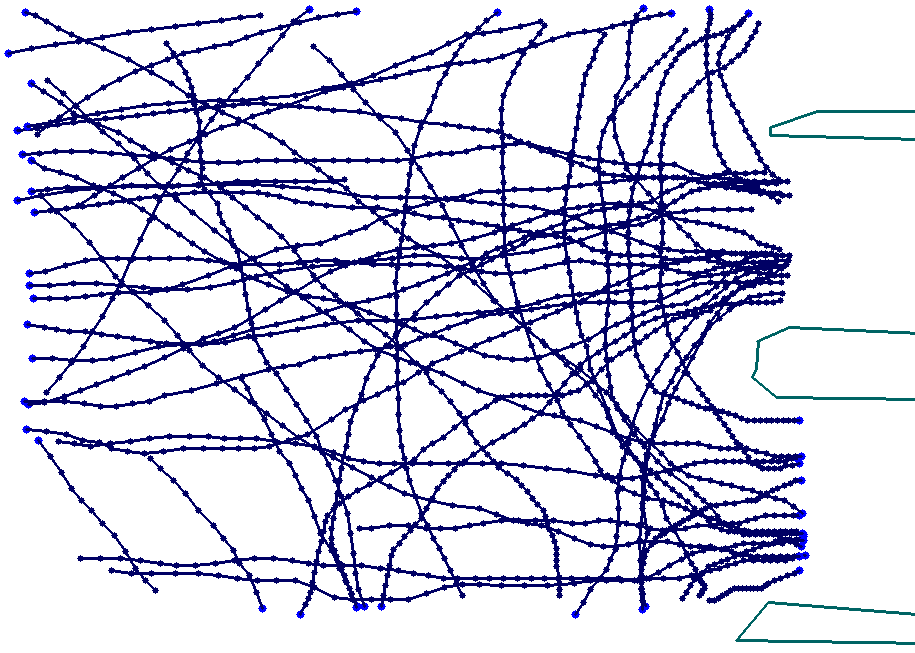}
    \caption{Video tracking used to extract the trajectories of pedestrians from video recordings close to
two escalators.
Left: Illustration of the tracking of pedestrian heads. Right: Resulting trajectories after being transformed onto the two-dimensional plane.}  
\label{Fig2}
\end{center} \end{figure}

We have made several video recordings of pedestrian crowds in different natural environments
in Budapest (Hungary) and Stuttgart (Germany) from the top (see Fig.~\ref{Fig1}). The dimensions of
the recorded areas was known, and the floor tiling or environment provided something like a
``coordinate system''. The heads were automatically determined by seaching for round moving structures,
and the accuracy of tracking was improved by comparing actual with linearly extrapolated positions
(so it would not happen so easily that the algorithm interchanged or ``lost'' closeby pedestrians). 
The trajectories of the heads were then projected on two-dimensional space in a way correcting for
distortion by the camera perspective. A representative plot of the resulting trajectories 
is shown in Fig.~\ref{Fig2}.
However, it should be noted that extracting trajectory data from pedestrians is nothing new. It 
has been done in the past with infra-red sensors \cite{KerridgeEmpirical} or video recordings \cite{HoogendoornEmpirical,Teknomo}.

\subsection{Parameter Optimization by an Evolutionary Algorithm}

For model calibration, we have used a hybrid method fusing empirical trajectory data
and microscopic simulation data of pedestrian movement in space. 
To each tracked pedestrian, we have assigned a virtual pedestrian 
in the simulation domain. We have then started a simulation for $T=1.5$ seconds, in which one pedestrian 
$\alpha$ was moved according to a simulation of the social force model, while the others were 
moved exactly according to the trajectories extracted from the videos.
This procedure was performed for all pedestrians $\alpha$
and for several different starting times $t$, using a fixed parameter set for the social force model.
\par
Each simulation run was performed according to the following scheme:
\begin{enumerate}
	\item Define a starting point and calculate the state 
(position $\vec{r}_{\alpha}$, velocity $\vec{v}_{\alpha}$, and acceleration 
$\vec{a}_{\alpha} = d\vec{v}_\alpha/dt$) for each pedestrian $\alpha$.
	\item Assign a desired speed $v_\alpha^0$ to each pedestrian. In our simulations,
we have specified it by the maximum speed $v_\alpha(t) = \|\vec{v}_\alpha(t)\|$ during the 
pedestrian's tracking time, which is sufficiently accurate if the overall pedestrian density is not
too high and the desired speed is constant in time. 
	\item Assign a desired goal point for each pedestrian. We have assumed is would
correspond to the point at the end of the trajectory.
	\item Given the tracked motion of the surrounding pedestrians $\beta$,
simulate the trajectory of pedestrian $\alpha$ based on 
the social force model during $T$ seconds of time, starting at the actual location
$\vec{r}_\alpha(t)$.
\end{enumerate}
After each run, we determined the relative distance error
\begin{equation}
\frac{\|\vec{r}_\alpha^{\rm simulated}(t+T) - \vec{r}_\alpha^{\rm tracked}(t+T)\|}
 {\|\vec{r}_\alpha^{\rm tracked}(t+T) - \vec{r}_\alpha^{\rm tracked}(t)\|} \, .
\end{equation}
After averaging the relative distance errors over the pedestrians $\alpha$ and starting times $t$, 
the negative value of the result
was taken as {\em fitness} of the parameter set used in the pedestrian simulation.\footnote{Note that
we have actually averaged over the 30\% central values of the relative distance errors only in order to remove
outliers due to untypical, unreasonable, or exceptional pedestrian behaviors.}
Hence, the best possible value
of the fitness was 0, but any deviation from the real pedestrian trajectories would imply negative fitness
values.
\par
Moreover, with the fitness measure outlined above, we have used a simple evolutionary 
algorithm to obtain the parameter set with the highest fitness value, which was $-0.39$.
This value reflects in some sense the stochasticity and/or heterogeneity of individual behaviors.
For comparison, we also made simulations with $A=0$, i.e. without any social forces, assuming that
persons will never change their velocity. This measurement resulted in a fitness value of $-0.66$.


\subsection{Improved Specifications of the Social Force Model}

\subsubsection{Velocity-Dependent Interaction Forces}

Interactions among pedestrians are actually more complicated than suggested above. For example,
it is know that the angle $\delta_{\alpha\beta}$ matters, at which two pedestrians $\alpha$ and $\beta$
approach each other (which is given by $\cos(\delta_{\alpha\beta}) =
\vec{e}_\alpha\cdot \vec{e}_\beta$). Apart from this, the step sizes and, therefore, the speeds matter as well.
In the following, we will shortly discuss two anisotropic models of pedestrian interactions:
\par
\begin{figure}[htb] 
\begin{center}
    \includegraphics[width=0.5\textwidth]{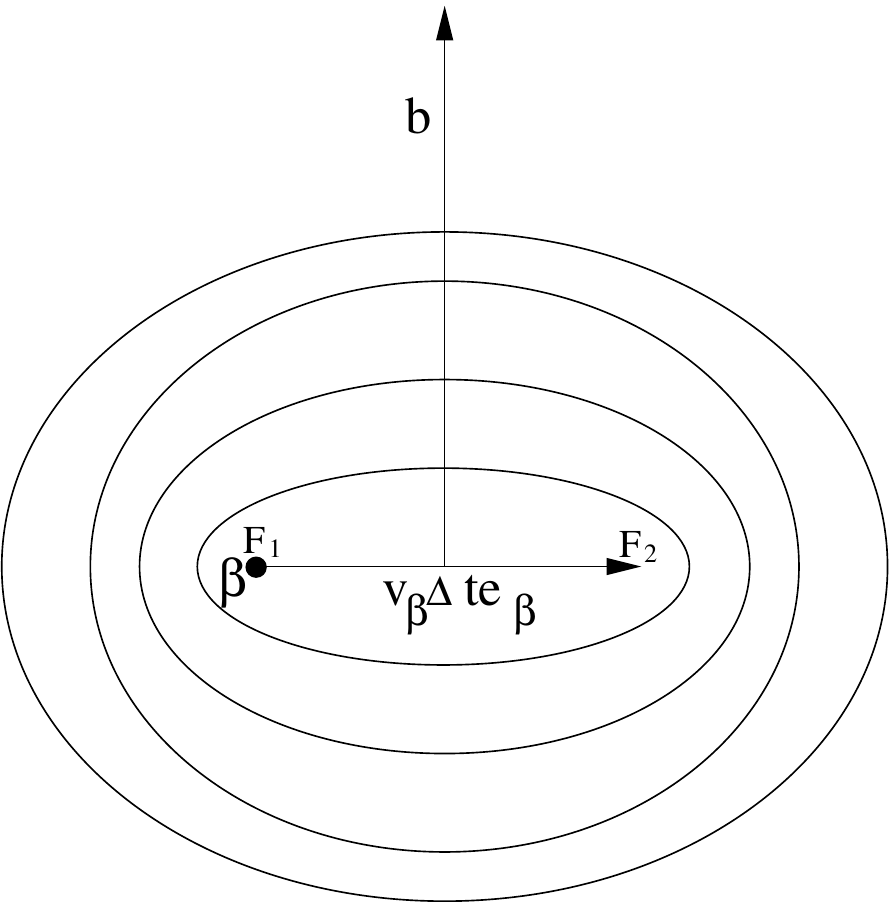}
    \caption[]{Illustration of the elliptical specification of pedestrian interaction forces.}
\label{ell1}
\end{center} 
\end{figure}

{\bf Elliptical specification I:} In Ref.~\cite{HelbMoln1995}, a generalization of Eq.~(\ref{g}) was
formulated, which assumed that the repulsive potential 
\begin{equation}
V_{\alpha\beta}(b)=AB \, \mbox{e}^{-b_{\alpha\beta}/B}
\end{equation}
is an exponentially decreasing function of $b$ with 
equipotential lines having the form of an ellipse directed into the direction of 
motion as shown in Fig.~\ref{ell1}. The semi-minor axis $b_{\alpha\beta}$ was determined by 
\begin{equation}
2b_{\alpha\beta} =\sqrt{(\|\vec{d}_{\alpha\beta}\| + \|\vec{d}_{\alpha\beta}-v_\beta\,\Delta t\,\vec{e}_{\beta}\|)^2 
- (v_{\beta}\, \Delta t)^2} 
\label{one}
\end{equation}
in order to take into account the length $v_\beta\,\Delta t$ of the stride (step size) of pedestrian $\beta$,
where $v_\beta = \|\vec{v}_\beta\|$. 
The reason for this specification was that pedestrians require space for movement, 
which is taken into account by other pedestrians. 
\par
The repulsive force is related to the repulsive potential via
\begin{equation}
\vec{g}_{\alpha \beta}(\vec{d}_{\alpha \beta})=-\vec{\nabla}_{\vec{d}_{\alpha \beta}} V_{\alpha \beta}(b_{\alpha\beta}) 
 = - \frac{dV_{\alpha \beta}(b_{\alpha\beta})}{db_{\alpha\beta}} \vec{\nabla}_{\vec{d}_{\alpha \beta}} b_{\alpha\beta}(\vec{d}_{\alpha\beta})\, .
\end{equation}
Considering the chain rule, $\|\vec{z}\| = \sqrt{\vec{z}^2}$, and $\vec{\nabla}_{\vec{z}}
\|\vec{z}\| = \vec{z}/\sqrt{\vec{z}^2} = \widehat{\vec{z}}$, this leads to the explicit formula
\begin{equation}
 \vec{g}_{\alpha \beta}(\vec{d}_{\alpha \beta}) = A \mbox{e}^{-b_{\alpha\beta}/B} \cdot
 \frac{\|\vec{d}_{\alpha\beta}\| + \|\vec{d}_{\alpha\beta}-\vec{y}_{\alpha\beta}\| }{2b_{\alpha\beta}}
\cdot \frac{1}{2} \left( \frac{\vec{d}_{\alpha\beta}}{\|\vec{d}_{\alpha\beta}\|} 
 + \frac{\vec{d}_{\alpha\beta}-\vec{y}_{\alpha\beta}}{\|\vec{d}_{\alpha\beta}-\vec{y}_{\alpha\beta}\|}\right)  
\end{equation}
with $\vec{y}_{\alpha\beta} = v_\beta\,\Delta t\,\vec{e}_{\beta}$.
For $\Delta t = 0$, we regain the expression of Eq.~(\ref{g}).
\par
{\bf Elliptical specification II:} Recently, a variant of this approach has been 
proposed \cite{DiplomaThesisShukla}, assuming
\begin{equation}
2b:=\sqrt{(\|\vec{d}_{\alpha \beta}\| + \|\vec{d}_{\alpha\beta}-(\vec{v}_\beta - \vec{v}_\alpha)
\Delta t \|)^2 - [(\vec{v}_\beta - \vec{v}_\alpha)\Delta t]^2} \, .
\label{two}
\end{equation}
The special feature of this approach is its symmetrical treatment of both pedestrians $\alpha$
and $\beta$.
\par
In order to compare the elliptical specifications with the circular one, 
let us make the following thought experiments:
\begin{itemize}
\item{\it Scenario 1:} 
Consider a pedestrian walking towards a wall with a given speed $v_\alpha$, but
four different orientations $\vec{e}_\alpha^{\,k}$ (with $k\in\{1,2,3,4\}$),
as shown in Fig. \ref{four}. This scenario can
be treated analogously to the movement relative to a 
standing pedestrian $\beta$,  which implies $\vec{v}_\beta = \vec{0}$.
\begin{figure}[htb] 
\begin{center}
    \includegraphics[width=0.9\textwidth]{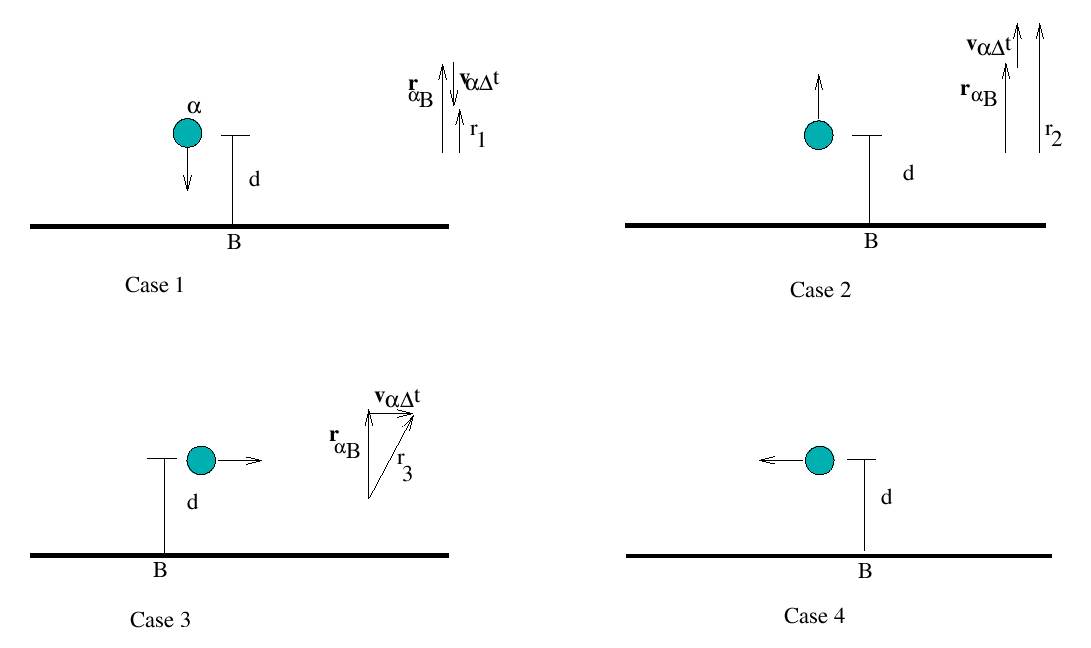}
\caption[]{Illustration of scenario 1, in which pedestrians are walking at a given speed
in different directions close to a wall.}
\label{four}
\end{center} 
\end{figure}
Then, in case II the repulsive force is a monotonously decreasing function of $b_{\alpha \beta}$ given by
\begin{equation}
 2b_{\alpha\beta}=\sqrt{(\|\vec{d}_{\alpha\beta}\| 
+ \|\vec{d}_{\alpha \beta}+v_\alpha\Delta t\vec{e}_{\alpha}\|)^2 - (v_{\alpha}\Delta t)^2} 
\end{equation}
according to Eq.~(\ref{two}). 
For all four cases, the values of $d:=\|\vec{d}_{\alpha \beta}\|$ and $v_{\alpha}\Delta t$ 
(the step size of pedestrian $\alpha$) are the same, but the values of 
$d_\alpha^k := \|\vec{d}_{\alpha \beta}+v_\alpha\Delta t\vec{e}_{\alpha}^k\|$ are different. 
We have $d_\alpha^1<d_\alpha^3=d_\alpha^4<d_\alpha^2$, so that we find  
$F_\alpha^1>F_\alpha^3=F_\alpha^4>F_\alpha^2$ for the magnitudes of the repulsive
forces triggered by the wall (as the direction of the forces is perpendicular to the wall in all four cases.)
This agrees well with experience, i.e. 
the anisotropic and orientation behavior of pedestrians are 
realistically reproduced by the elliptical force specification II. In contrast, the elliptical model I
implies $b_{\alpha\beta} = d_{\alpha\beta}$ according to Eq.~(\ref{one}) and predicts the same
force in all four cases, as does the circular model.

\item{\it Scenario 2:} 
Consider single pedestrians $\alpha$ walking towards a wall, but with 2 different 
velocities $(v_\alpha^1<v_\alpha^2)$ as shown in Fig. \ref{speeds}. Obviously, the
elliptical specification II predicts that the pedestrian $\alpha$ with the higher speed would decelerate earlier, 
as expected. In contrast, the elliptical model I and the circular interaction model
are insensitive to the own speed $v_\alpha$.
\begin{figure}[htb] 
\begin{center}
    \includegraphics[width=0.9\textwidth]{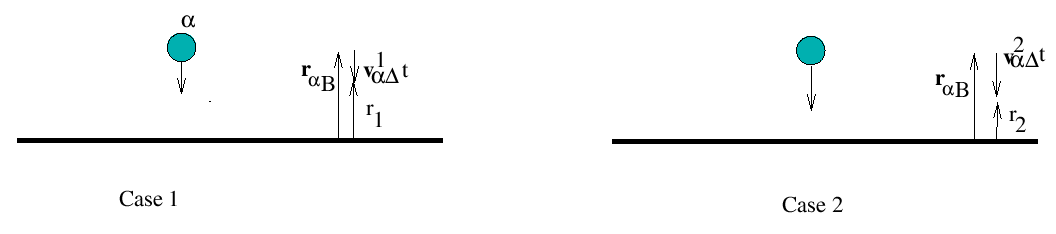}
\caption[]{Illustration of scenario 2, in which pedestrians are 
walking with different speeds towards a wall.} 
\label{speeds}
\end{center} 
\end{figure}
\item{\it Scenario 3:} 
Consider the case of two pedestrians walking towards each other with different velocities $v_{\alpha}>v_{\beta}$,
as shown in Fig.~\ref{towards}.
\begin{figure}[htb] 
\begin{center}
    \includegraphics[width=0.3\textwidth]{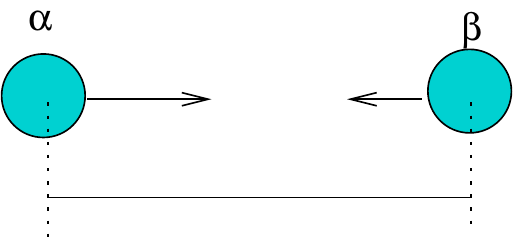}
\caption[]{Illustration of scenario 3, in which pedestrian are walking towards each 
other with different velocities.} 
\label{towards}
\end{center} 
\end{figure}
While the circular interaction force does not imply any velocity-dependence of pedestrian behavior,
the elliptical formulations do, which is certainly more realistic. The first elliptical model assumes
a velocity-dependence of pedestrian interactions, but 
no sensitivity with respect to the {\em own} speed. Therefore, 
specification II is expected to be more realistic.
\end{itemize}
For the above reasons, we will not consider the elliptical specification I anymore,
but focus on the elliptical model II.

\begin{figure}[htb] \begin{center}
    \includegraphics[width=0.75\textwidth]{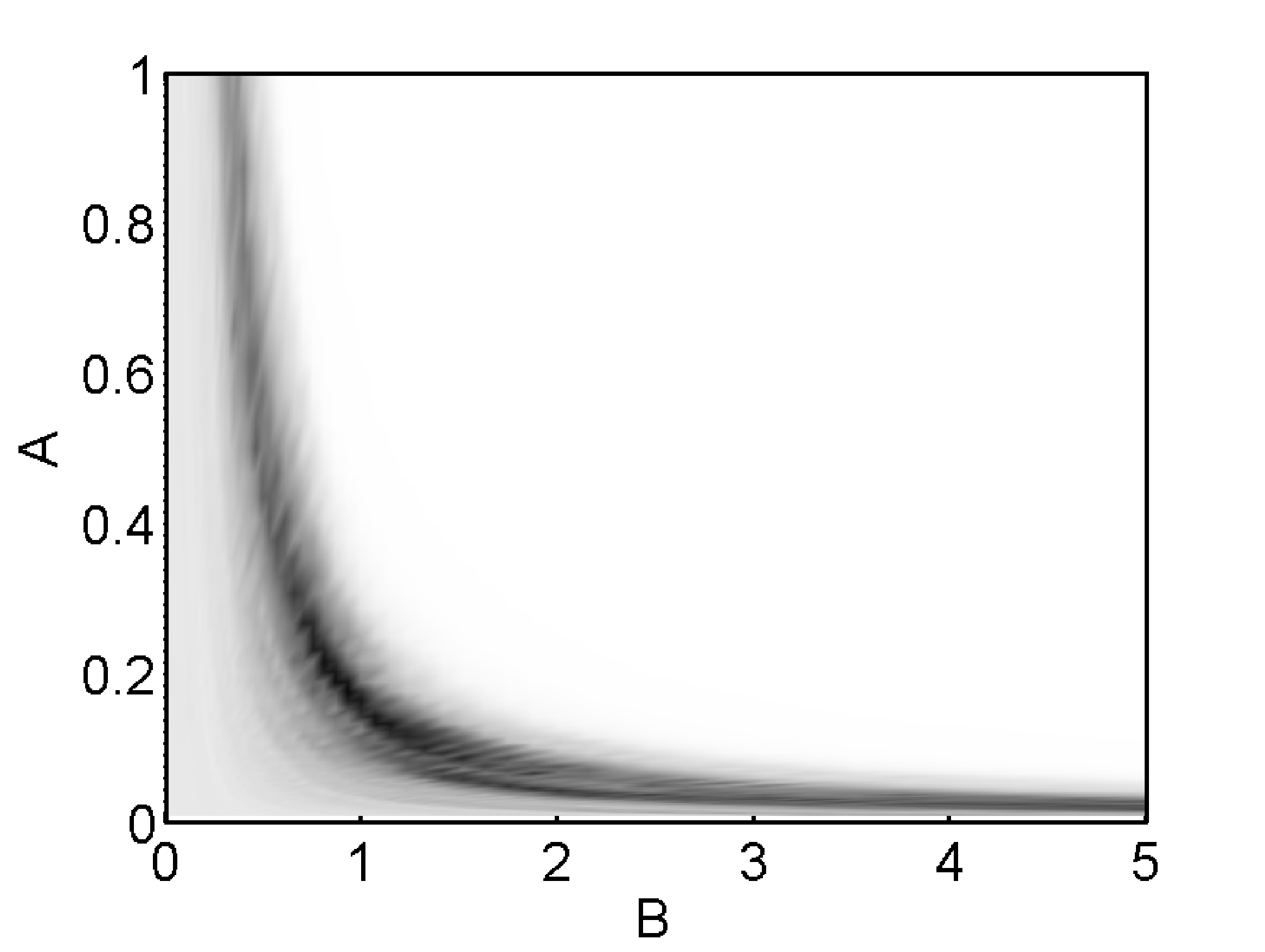}
      \caption{Goodness of fit for different parameter values $A$ and $B$, when pedestrian trajectories are simulated with the
social force model and the the force specification (\ref{g}). The darker the shade, the better the fitness is, for the corresponding combination of $A$ and $B$ values.}
\label{Fig_3} \end{center} \end{figure}

Figure~\ref{Fig_3} represents the resulting fitness values as a function of the choice of
the interaction strength $A$ and the interaction range $B$, assuming $w(\varphi) \equiv 1$.
\par\begin{table}[htbp]
\caption[]{Optimal parameter values resulting from our evolutionary parameter optimization 
for three different specifications of the interaction forces between pedestrians (see main text)
The parameter calibration was based on the 3 video recordings mentioned in Fig.~\ref{Fig1}
and the fundamental diagram of Weidmann, see Eq.~(\ref{Weidmann}).}
\label{Tab}
\begin{center}
\begin{tabular}{l|ccc|r}
\hline
Model & A & B & $\lambda$ & Fitness \\
\hline
Extrapolation & 0 & -- & -- & -0.66 \\
Circular & 0.42 $\pm$ 0.26 & 1.65 $\pm$ 1.01 & 0.12 $\pm$ 0.07 & -0.60 \\
Elliptical I & 0.11 $\pm$ 0.01 & 1.19 $\pm$ 0.45 & 0.16 $\pm$ 0.04 & -0.59 \\
Elliptical II & 0.04 $\pm$ 0.01 & 3.22 $\pm$ 0.67 & 0.06 $\pm$ 0.04 & -0.39 \\
Circular & 0.11 $\pm$ 0.06 & 0.84 $\pm$ 0.63 & 1 & -0.65 \\
Elliptical I & 1.52 $\pm$ 1.65 & 0.21 $\pm$ 0.08 & 1 & -0.67 \\
Elliptical II & 4.30 $\pm$ 3.91 & 1.07 $\pm$ 1.35 & 1 & -0.47 \\
\hline
\end{tabular}
\end{center}
\end{table}
One result of our parameter optimization was that, for each video, there was actually a broad range
of parameter combinations of $A$ and $B$ which performed almost equally well, while
the optimal value of the anisotropy parameter was $\lambda = 0.11 \pm 0.07$. 
This allowed us to apply additional
goal functions in our optimization. We used this fact to determine among the
best performing parameter values such parameter combinations, which performed well for 
{\em all three} video recordings, using a fitness function which weighted the fitness reach in
each single video equally (i.e. with a factor 1/3). 

\begin{figure}[htb] \begin{center}
    \includegraphics[width=0.7\textwidth,angle=-90]{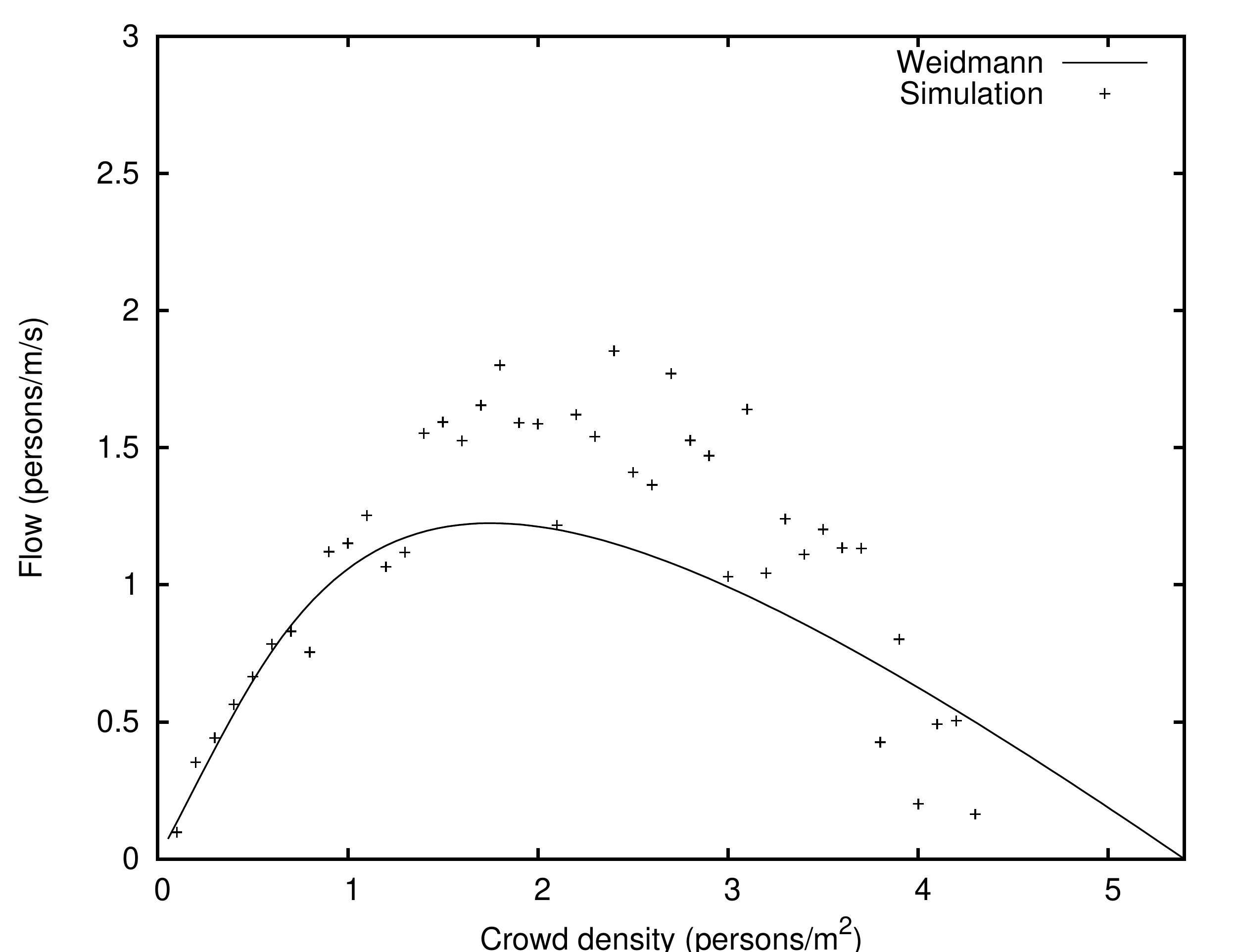} 
    \caption[]{Comparison of the fundamental diagram for pedestrian flows 
by Weidmann, see Eq.~(\ref{Weidmann}) with the fundamental diagrams resulting 
in simulations with the social force model, Elliptical II, for the parameter combinations
listed in Tab.~\ref{Tab}.}
\label{fundcomp}
\end{center} \end{figure}

The resulting parameters are listed in Table~\ref{Tab}.
The parameters corresponding to the best fitness value, i.e. the elliptical specification II with $A=0.04, B=3.22, \lambda=0.06$, were used in a microscopic 
simulation with periodic foundary conditions, and the resulting fundamental diagram, is shown in Fig.~\ref{fundcomp}.
The solid line is the Weidmann curve, Eq.~(\ref{Weidmann}), 
Note that, with additional fine tuning, it will be possible to make an even better fit to the Weidmann fundamental diagram, but here
we rather just show that extracting parameters from videos and perform a microscopic simulation will 
correspond to dynamics close to the empirical flow-density relation.

\par

\begin{figure}[htb] \begin{center}
    \includegraphics[width=0.45\textwidth]{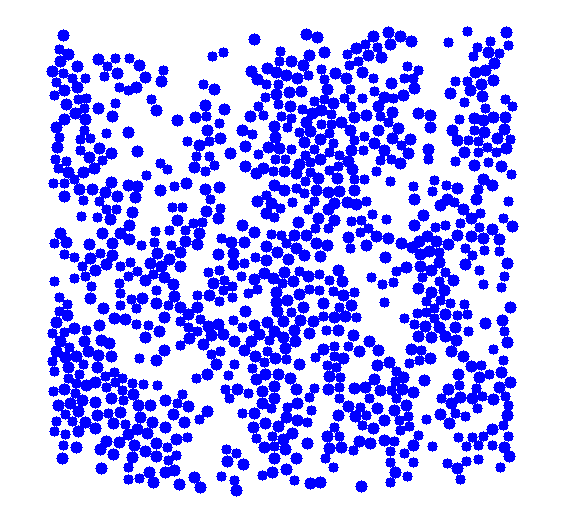}
    \includegraphics[width=0.45\textwidth]{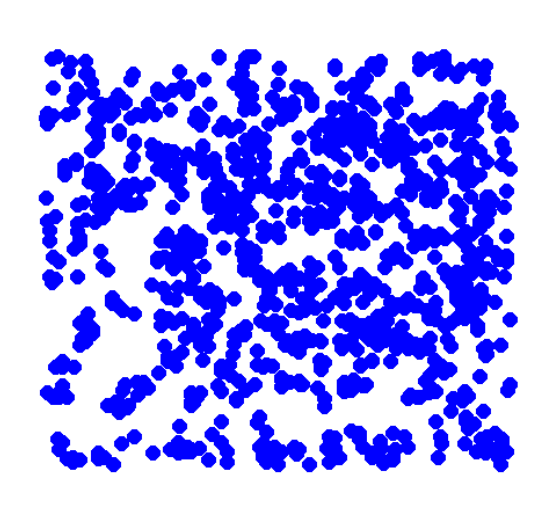}
      \caption{Another good feature of the elliptical model II is that it produces a non-regular density distribution, which is comparable to how persons distrubute in dense crowds, in reality. Left: A snapshot from a simulation with the elliptical model II. Right: A snapshot from reality, where persons are identified from a picture of a dense crowd. In both cases, the persons are walking from left to right.}
\label{Fig_xxx} \end{center} \end{figure}

%
%
We would, however, like to mention that further velocity-dependent
specifications of pedestrian interaction forces were proposed in the past, for example, by 
Moln\'{a}r Ref.~\cite{PhDthesisPeterMolnar,Verkehrsdynamik}, Okazaki \cite{MagneticForceModel}, 
Hoogendoorn \cite{Hoog2002b}, and Yu \cite{YuPRE}.
\par
Although, from a theoretical perspective, velocity-dependent force specifications
are more promising than the circular interaction force introduced in the beginning, 
it turns out that the prediction of pedestrian motion does not become
a lot better by including additional, speed-dependent parameters. This is illustrated
by the fitness values in Table \ref{Tab}. Obviously, the deviation of 
the fitness value from 0 is mainly caused by the 
heterogeneity of pedestrian behaviors. A similar conclusion has been drawn from 
studies specifying behavioral parameters in 
different car-following models \cite{Wagner}.

\subsubsection{Angular Dependence of Pedestrian Interactions}

\begin{figure}[htb] \begin{center}
    \includegraphics[width=0.75\textwidth]{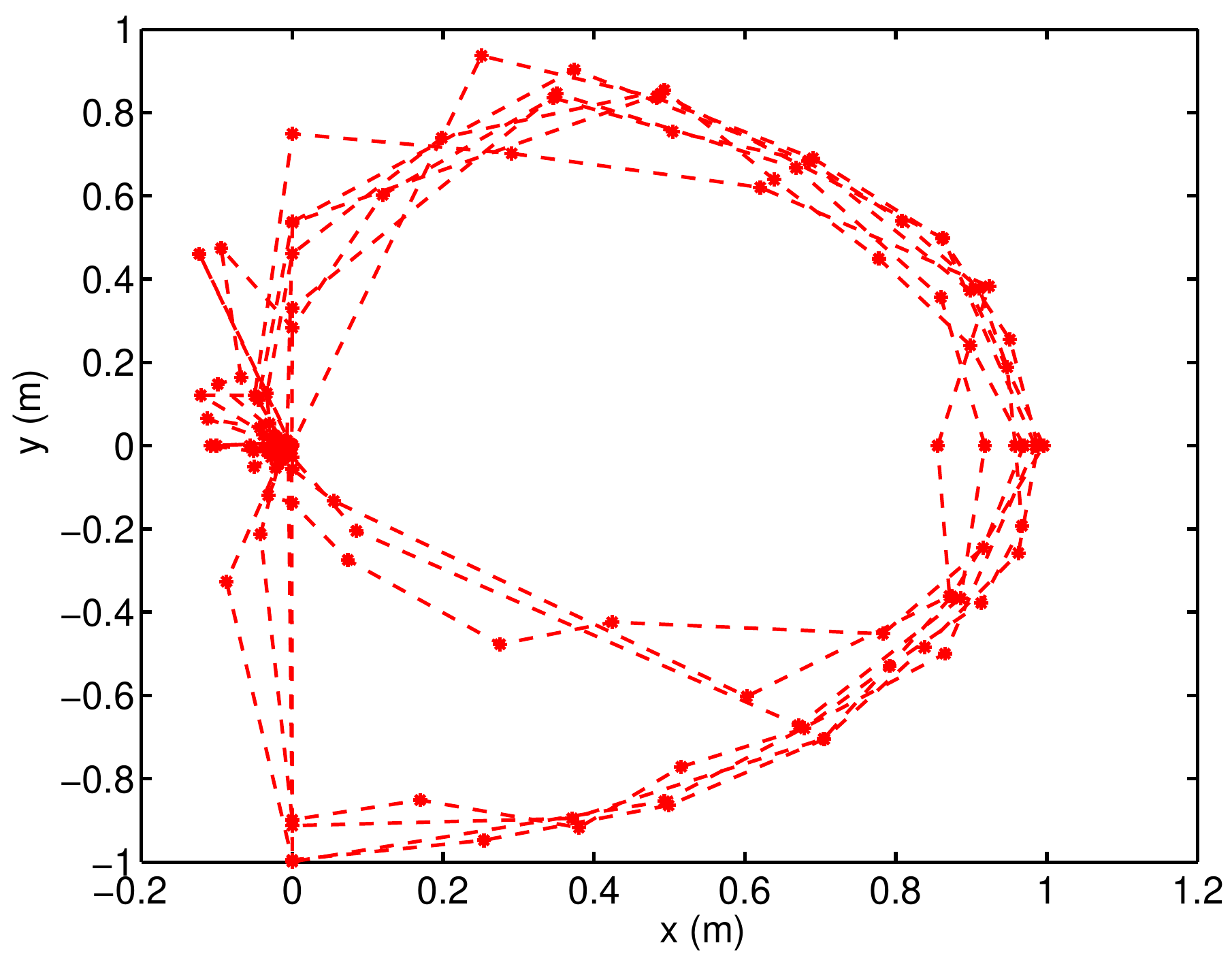}
    \caption{Angular dependence of the influence of other
pedestrians. The direction along the positive $x$ axis corresponds to the
walking direction of the pedestrian, $y$ to the perpendicular direction.}
\label{Fig4} \end{center} \end{figure}

Our evolutionary fitting method can be also used to determine interaction laws without
prespecified interaction functions. For example, one can 
determine the angular dependence of pedestrian interaction 
directly. In order to do this, we have chosen an approach using a polygon with $n$ edges
in the directions $\varphi \in 2 \pi i/n, i \in [0, n-1]$. The distance of the
curve in direction $\varphi_i$ from the origin represents the (relative) angular
interaction strength $w(\varphi_i)$ with a maximum value of 1.
To get some smoothness, linear interpolation has been applied between each of
the edges.
Figure \ref{Fig4} shows the angular dependence of the interaction strength determined
from pedestrian trajectories. Apparently, pedestrians are only sensitive to what happens inside an $180$
degree angle in front of him or her, which roughly corresponds to the visually perceived area.

\subsubsection{Distance Dependence of Pedestrian Interactions}

\begin{figure} 
\begin{center}
	\includegraphics[width=0.7\textwidth]{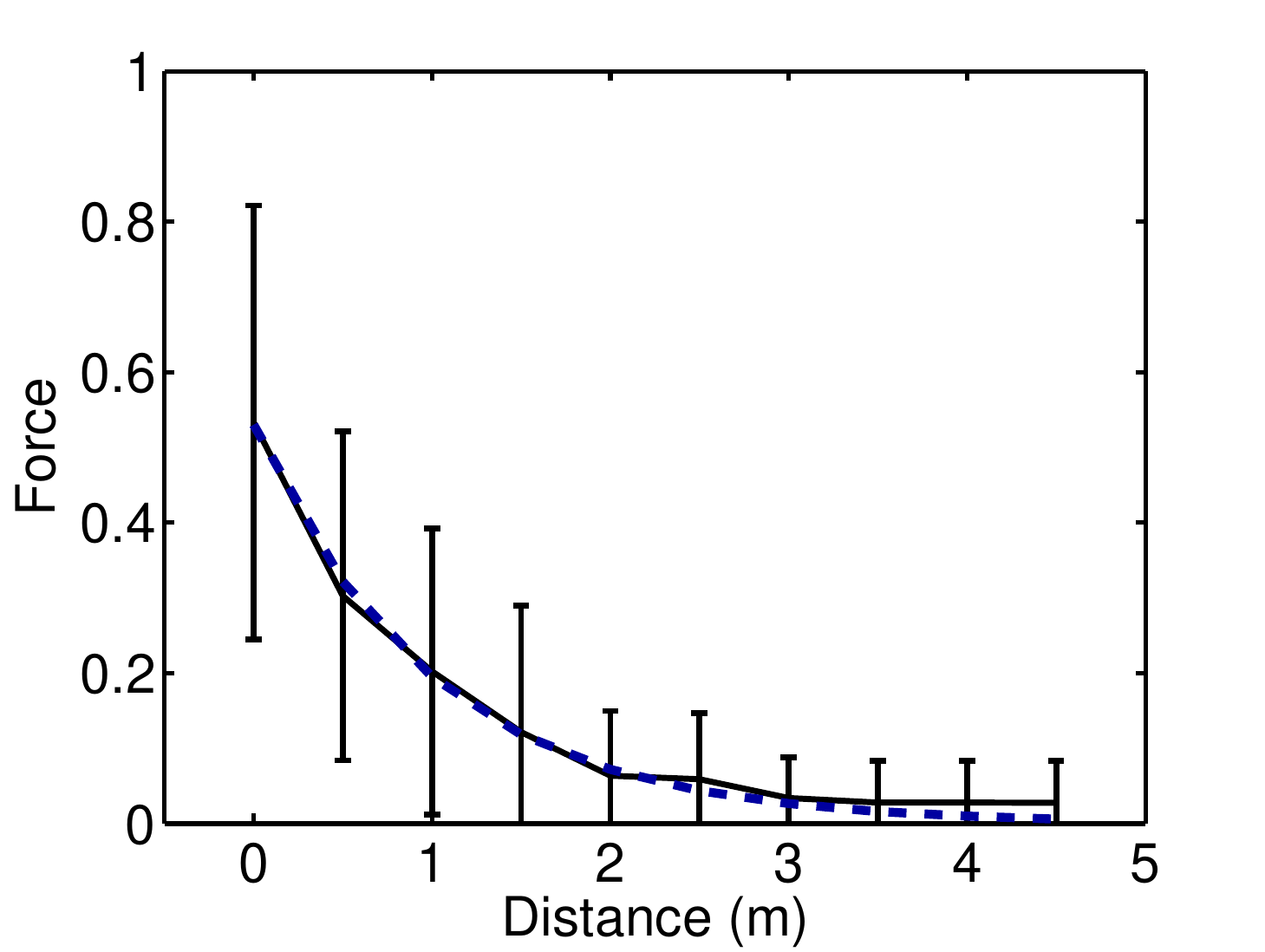}
\end{center}
\caption{Empirically determined distance dependence of the interaction force between pedestrians.
An exponential decay fits the empirical data quite well. The dashed fit curve is made with
Eq. \ref{g} and $A=0.53, B=1.0$. From the simulation with the circular model the resulting parameters
$A=0.42, B=1.62$ are quite close.}
\label{expo}
\end{figure}

In a similar way, we have determined the distance dependence of pedestrian interactions without
a pre-specified function. For this, we have adjusted the values of the force at given distances
$d_k = k d_1$ (with $k\in \{1,2,3,...\}$) in an evolutionary way. The resulting fit curve is presented 
in Fig.~\ref{expo}. According to our results, the empirical dependence of the force with distance can be 
well fitted by an exponential decay, as assumed before. Note that similar studies have been carried
out for vehicle interactions, recently 
\cite{LetztesPhysicaAPapermitTreiber}. However, the situation for pedestrians is more difficult due
to their two-dimensional motion. Therefore, our evolutionary fitting procedure seems to be superior to the
Fokker-Planck approach (or Random Matrix Theory approach), which was used to calibrate vehicle interactions.

\section{Large-Scale Pedestrian and Evacuation Simulations}\label{Sec3}

In the following section, we will present some examples for large-scale pedestrian simulations with the
social force model. For simplicity and computational speed, we have assumed circular pedestrian interactions,
here. Compared to previous implementations of the model, numerical efficiency
has been increased a lot, for example by cutting off interation forces at a certain, large enough
distance (around $5$ m)
and book-keeping of neighboring pedestrians. In this way, the simulation of 30,000 pedestrians or more
in real-time is no problem. Larger crowds have been simulated off-line or using parallel computing.
\par
\begin{figure}[htbp]
\begin{center}
\includegraphics[width=11cm]{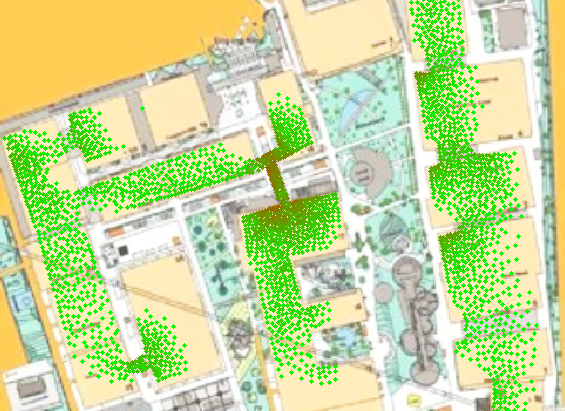} 
\end{center}
\caption{Snapshot of an evacuation simulation of a convention or exposition hall.}
\label{EXPO}
\end{figure}

For example, a simulation have been performed for an evacuation of two decks 
of a ship connected by a staircase. The underlying study has shown
that the model is also suitable for treating different personalities (the different
characteristic behaviors in emergency situations),
complex geometries, and three-dimensional interactions \cite{werner}.
A simulation \cite{Hoog2002b} have been performed for multi-destination flows at Schiphol airport in the Netherlands, 
while \cite{Quinn} have simulated the evacuation of an exhibition area, both with the social force model. 
For a similar simulation see Fig.~\ref{EXPO}.
\par
\begin{figure}[!ht]
\begin{center}
\includegraphics[width=10cm]{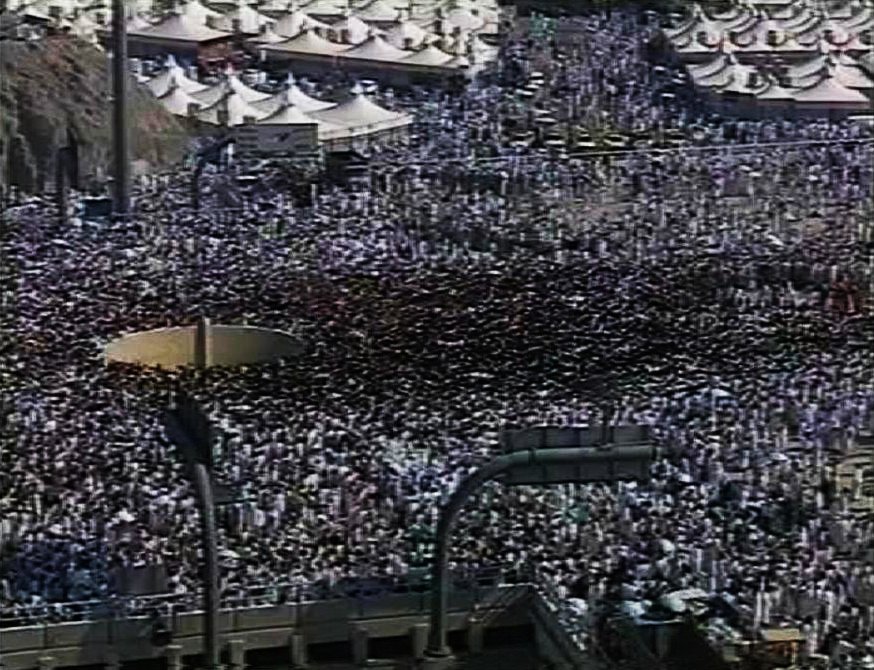}\\
\includegraphics[width=10cm]{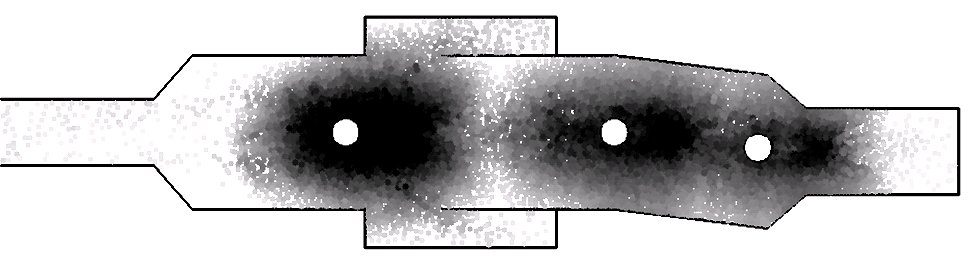} 
\end{center}
\caption[]{Top: Simulation of the Jamarat bridge 
which has to cope each year with millions of pilgrims within a few days. 
The pilgrims need to get close to the circular area in order to throw 
stones at the Jamarah (the pillar in the center representing
the devil). This leads to extreme pedestrian densities behind 
the pillars (dark red and light yellow areas) which have led to fatalities in the past. 
Note that the recent replacement of the circular Jamarahs by elliptical ones has lead to considerable
improvements in the capacity and safety of the stoning process. 
Bottom: Photograph of the (dark)
compression area behind a circular Jamarah.
\label{jamarat}}  
\end{figure}

Our model has also been used to simulate pedestrian flows on the Jamarat Bridge,
which has to cope with 3 million pilgrims within a single day 
(AlGadhi and Mahmassani, 1990). Figure~\ref{jamarat} illustrates the scenario for the old Jamarat
bridge. It shows the development of zones of high density and pressure behind the previous 
circular basins. In the middle of each of the three basins is a pillar symbolizing the devil. 
These pillars are to be stoned with
seven pebbles each, which takes time and delays pedestrian motion. This, in turn, can cause
extreme crowding. Note, however, that the situation around the pillars has significantly improved 
since the circular stoning areas were replaced by elliptical ones. 
\par
\begin{figure}[!htbp]
\begin{center}
\includegraphics[width=11.5cm]{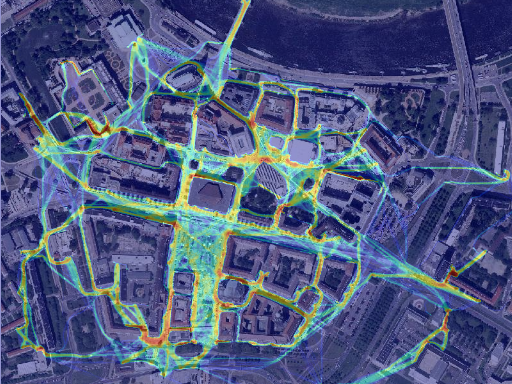} 
\end{center}
\caption[]{Densities of pedestrian streams in a simulation of the city center of
Dresden, Germany. Our simulation handles 30.000 pedestrians.\label{urban}}
\end{figure}

As indicated before, the implementation of our pedestrian microsimulator has been optimized for
efficiency and scalability in terms of computational speed and memory. It is no problem to simulate
more than 30,000 pedestrians in real-time on a normal PC. Parallel computing even allows one to simulate
more than 100,000 pedestrians in real-time. This is suitable for the simulation of
large scenarios such as pedestrian flows in extended urban areas (see Fig.~\ref{urban})
or during mass events such as street parades or carnivals \cite{Batty2}.
\par
In order to be suitable for scenarios of this complexity, we have developed 
a powerful and flexible pedestrian simulator called ``UNIVERSE'', in which we have
implemented certain additional features complementing the social force model:
\begin{enumerate}
\item a powerful scenario generation module based on XML,
\item path finding and route choice modules 
to describe the selection of destinations,
including spontaneous stops and impluse shopping,
\item calibration modules to adapt simulations to real measurements,
\item a statistics module evaluating
various performance measures (such as densities, velocities, flow efficiency, pressure, overall
evacuation time, etc.) to assess and compare different scenarios,
\item an optimization module based on evolutionary algorithms which allows to
improve the geometrical boundary conditions of pedestrian facilities,
\item various visualization modes for the illustration of the simulation results, both on- and
off-line, and
\item a functionality reflecting density-dependent herding effects.
\end{enumerate}
\section{Summary and Outlook}\label{Sec4}

Tracking algorithms have become a powerful tool for empirical data evaluation, recently \cite{HoogendoornEmpirical,Teknomo,KerridgeEmpirical}.
In our study, we have further developed and
applied this tool to determine trajectories of many interacting pedestrians. These
trajectories were used to calibrate a microscopic pedestrian model, namely the social force model. 
For this, we have simulated single
pedestrians, given the measured motion of the other pedestrians. The difference between simulated
and measured pedestrian behavior has then been minimized through evolutionary variation of the 
model parameters. However, it was also possible to determine
the distance- and angle-dependencies of the interactions directly. In this way, the previously
assumed exponentially decay of the interaction strength with distance was empirically confirmed.
\par
It turned out that the simplest specification of the interaction forces in the social force 
model almost performed as well as velocity-dependent specifications with more parameters. 
We have, therefore, used the circular specification for our large-scale simulations, which is in favor
of computational speed. With some simplifications for the sake of numerical performance,
it is no problem to simulate scenarios with 30,000 pedestrians on-line on a normal
PC. Larger crowds can be simulated off-line or using parallel computing.
This allows one to study pedestrian zones, convention halls, shopping malls, airports, railway stations,
city festivals, etc. There are, by the way, also special algorithms for the evolutionary optimization of 
highly frequented areas \cite{OptimizationPaperbyAnders}. Hence, pedestrian
simulations are now a useful tool for the planning of pedestrian facilities and mass events,
if properly calibrated. 
\par
Note, however, that that measurement and calibration results should be transfered
from one place to another only with care, as the maximum flow and density
values are largely influenced by the body size distribution.
Other factors may play a role as well, for example
the age distribution, the fraction of handicapped people, the carrying of luggage, alcohol consumption,
cultural habits, use of mobile phones, and the situational context \cite{MilinskiBuch}  
(e.g. business, leisure, sports, or religious activities). 
All of this should be taken into account in the dimensioning
of pedestrian facilities, safety (evacuation) analyses, and the calibration of computer models.

\subsection*{Acknowledgements}

The authors would like to thank Peter Felten for his work with the video material and
the German Research Foundation (DFG project He 2789/7-1) for financial support.
Further on, the authors are thankful to Serge Hoogendoorn for providing empirical data, used to verify the results, shown in figures \ref{Fig4} and \ref{expo}.

\end{document}